\documentclass[runningheads]{llncs}
\usepackage[T1]{fontenc}
\usepackage{graphicx}

\usepackage{cite}
\usepackage{url}
\usepackage{comment}
\usepackage[hidelinks]{hyperref}

\usepackage[strict]{changepage}
\usepackage{colortbl}
\usepackage{framed}
\usepackage[acronym, shortcuts]{glossaries-extra}
\usepackage{multirow}
\usepackage{pifont}
\usepackage[moderate, leading=normal]{savetrees}
\usepackage{scalerel}
\usepackage{subcaption}
\usepackage{tikz}
\usepackage{bm}
\usepackage{soul} %
\usepackage{hyphenat} %

\glssetcategoryattribute{acronym}{nohyper}{true}
\setabbreviationstyle[acronym]{long-short}

\newacronym{ai}{AI}{Artificial Intelligence}
\newacronym{asr}{ASR}{Attack Success Rate}
\newacronym{ads}{ADS}{Anomaly Detection System}
\newacronym{can}{CAN}{Controller Area Network}
\newacronym{cnn}{CNN}{Convolutional Neural Network}
\newacronym{cpu}{CPU}{Central Processing Unit}
\newacronym{crc}{CRC}{Cyclic Redundancy Check}
\newacronym{ddos}{DDoS}{Distributed Denial of Service}
\newacronym{dnn}{DNN}{Deep Neural Network}
\newacronym{ecu}{ECU}{Electronic Control Unit}
\newacronym{ev}{EV}{Electric Vehicle}
\newacronym{far}{FAR}{False Acceptance Rate}
\newacronym{fas}{FAS}{Fingerprint-based Authentication Systems}
\newacronym{fgsm}{FGSM}{Fast Gradient Sign Method}
\newacronym{fn}{FN}{False Negative}
\newacronym{fp}{FP}{False Positive}
\newacronym{gan}{GAN}{Generative Adversarial Network}
\newacronym{gps}{GPS}{Global Positioning System}
\newacronym{gpu}{GPU}{Graphics Processing Unit}
\newacronym{gru}{GRU}{Gated Recurrent Unit}
\newacronym{ids}{IDS}{Intrusion Detection System}
\newacronym{imu}{IMU}{Inertial Measurement Unit}
\newacronym{lstm}{LSTM}{Long Short-Term Memory}
\newacronym{lstmfcn}{LSTM-FCN}{Long Short-Term Memory Fully Convolutional Network}
\newacronym{ml}{ML}{Machine Learning}
\newacronym{dl}{DL}{Deep Learning}
\newacronym{obd}{OBD}{On-Boad Diagnostic}
\newacronym{oem}{OEM}{Original Equipment Manufacturer}
\newacronym{rnn}{RNN}{Recurrent Neural Network}
\newacronym{shap}{SHAP}{SHapley Additive exPlanations}
\newacronym{svdd}{SVDD}{Support Vector Domain Description}
\newacronym{svm}{SVM}{Support Vector Machine}
\newacronym{tn}{TN}{True Negative}
\newacronym{tp}{TP}{True Positive}
\newacronym{tpm}{TPM}{Trusted Platform Module}
\newacronym{ubm}{UBM}{Universal Background Model}
\newacronym{ubi}{UBI}{Usage Based Insurance}
\newacronym{uds}{UDS}{Unified Diagnostic Services}
\newacronym{mitm}{MitM}{Man-in-the-Middle}
\newacronym{rpm}{RPM}{Revolutions Per Minute}
\newacronym{rf}{RF}{Random Forest}
\newacronym{gmm}{GMM}{Gaussian Mixture Model}
\newacronym{knn}{kNN}{k-Nearest Neighbors}
\newacronym{xai}{XAI}{Explainable Artificial Intelligence}
\newacronym{rl}{RL}{Reinforcement Learning}
\newacronym{sota}{SotA}{State-of-the-Art}

\newcommand*\halfcirc[1][1ex]{%
  \begin{tikzpicture}
  \draw[fill] (0,0)-- (90:#1) arc (90:270:#1) -- cycle ;
  \draw (0,0) circle (#1);
  \end{tikzpicture}}
\newcommand*\fullcirc[1][1ex]{\tikz\fill (0,0) circle (#1);}

\definecolor{formalshade}{rgb}{0.85,1,0.85}
\definecolor{darkblue}{rgb}{0.0,0.6,0.30}

\begin{document}
\title{When Authentication Is Not Enough: Breaking Behavior-Based Driver Authentication Systems}
\titlerunning{Breaking Behavior-Based Driver Authentication}
\author{
Emad Efatinasab\inst{1}\orcidID{0000-0003-3812-436X} \and
Francesco Marchiori\inst{1}\orcidID{0000-0001-5282-0965} \and
Denis Donadel\inst{2}\orcidID{0000-0002-7050-9369} \and
Alessandro Brighente\inst{1}\orcidID{0000-0001-6138-2995} \and
Mauro Conti\inst{1,3}\orcidID{0000-0002-3612-1934}
}
\authorrunning{Efatinasab et al.}
\institute{University of Padova, Padova, Italy\\
\email{\{emad.efatinasab,francesco.marchiori.4\}@phd.unipd.it, \{alessandro.brighente, mauro.conti\}@unipd.it}\\ \and
University of Verona, Verona, Italy\\
\email{denis.donadel@univr.it} \\ \and 
Örebro University, Örebro, Sweden
}
\maketitle              %
\begin{abstract}
Researchers extensively explored behavior-based driver authentication systems in vehicles.
Pushed by advances in Artificial Intelligence (AI), these systems employ powerful models to identify drivers based on unique biometric behaviors.
However, existing work prioritizes AI performance metrics, neglecting secure integration with real-world automotive environments and the threat of adversarial attacks that can fool the authentication system.

In this paper, we propose for the first evasion attacks against behavior-based driver authentication systems, allowing an attacker to impersonate the legitimate driver.
Our attacks exploit long-standing CAN bus weaknesses that allow the injection of forged frames without jeopardizing the attacker's safety while stealing the vehicle.
When legitimate data samples are available, we propose \textbf{SMARTCAN}, a safety-aware replay attack.
If the attacker can only use the authenticator as an oracle, we propose \textbf{GANCAN}, which trains a Generative Adversarial Network's generator using reinforcement learning on the authenticator's responses.
Our attacks achieve a success rate up to 100\% against all the considered models and, in the worst case, require 22 minutes to steal a vehicle.
Acknowledging our identified vulnerabilities, we discuss the requirements for a safe and effective deployment of these systems in real-world scenarios.

\end{abstract}

\section{Introduction}
\label{sec:introduction}

Traditional vehicle access methods, such as physical or wireless car keys, have long served as the primary means of driver authentication, linking drivers to their vehicles.
However, evolving technology and emerging security threats have exposed vulnerabilities in relying solely on car keys~\cite{garcia2016lock, francillon2011relay}.
Even with stronger cryptographic protections~\cite{garcia2016lock}, recent attacks exploiting direct access to the \ac{can} bus reveal ongoing weaknesses~\cite{miller2015remote, de2022canflict, kulandaivel2021cannon}. Smart keys, while more advanced, remain vulnerable to relay~\cite{francillon2011relay} and distance reduction attacks~\cite{leu2022ghost}. Biometric authentication is gaining traction~\cite{taha2023eyedrive}, but faces growing risks from advances in \ac{ai} and generative models, which can forge realistic biometric data and deceive even sophisticated systems~\cite{khanjani2021deep, stanciu2016effectiveness}.

In response to these enduring threats, researchers have proposed behavior\hyp{}based driver authentication~\cite{kang2019automobile, ravi2022driver, TSENG2023105571, xun2019automobile, ahmadi2021optimising, abu2020livedi} and identification~\cite{abdennour2021driver, azadani2020performance, chen2019driver, ezzini2018behind, marchegiani2018long, shin2025driver} 
systems to complement traditional methods.
These systems collect and analyze data, often from the \ac{can} bus, to either confirm that the driver is authorized (authentication) or identify the specific individual behind the wheel (identification) while the vehicle is operating.
To achieve this, they employ a range of models, from classical \ac{ml} approaches like \ac{svm} and \ac{rf} to advanced \ac{dl} methods. 
While the vehicle is in motion, these models continuously identify drivers by recognizing unique behavioral patterns, representing a shift in vehicle security with applications for tasks such as theft prevention, \ac{ubi} policies, and impaired driver detection~\cite{nyt_behavior, cura2020driver, meiring_review_2015}.

Despite growing research interest, existing behavior-based authentication and identification systems overlook key security concerns in practical deployment.
Furthermore, the current literature has neglected possible adversarial attacks on these systems, which can jeopardize their functionality.
Given the lack of a formalized system and threat model, the effects of \ac{ml} vulnerabilities in this scenario are unclear.
Challenges such as when and how to introduce adversarial inputs have yet to be explored at the time of writing.
To address these limitations, we identify the following research questions that outline critical gaps hindering the practical adoption of such systems.

\begin{itemize}
\item[\textbf{RQ1}]\label{rq1} 
Which threat vectors affect behavior-based driver authentication system deployed on a CAN bus?
\item[\textbf{RQ2}]\label{rq2} To what extent are \ac{sota} behavior based driver authentication system vulnerable?
\item[\textbf{RQ3}]\label{rq5} 
To what extent can these systems be used as a replacement to traditional authentication mechanisms?
\end{itemize}

\noindent{\textit{Contributions.}}
In this paper, we show that existing authentication systems are vulnerable to severe security issues that allow an attacker to impersonate the legitimate driver.
We developed the first two adversarial attacks against these systems, demonstrating their effectiveness against \ac{sota} models trained on real-world data. Then, to secure these authentication and identification systems, we provide an analysis of possible mitigations.
Table~\ref{tab:comparison} compares our system's contributions with the main works in the literature.
Our contributions can be summarized as follows.
\begin{itemize}
    \item We introduce \textbf{SMARTCAN} and \textbf{GANCAN}, the first attacks against behavior-based driver authentication and identification systems.
    They employ the attacker's knowledge---legitimate data or model responses, respectively\hspace{0pt}---and exploit evasion techniques to mimic legitimate users while injecting minimal spoofed \ac{can} frames to preserve driving functions and the driver's safety.

    \item We develop two new lightweight behavior-based driver authentication and identification systems whose requirements are compatible with commercial vehicle networks.
    Our architectures---a \ac{rf} and a single-layer \ac{gru}---are optimized for deployment in real-world conditions and outperform prior work, achieving up to 99.9\% accuracy.

    \item We evaluate our attacks on \ac{sota} and our lightweight models, achieving attack success rates up to 100\%, enabling an attacker to evade detection as an unauthorized driver in less than 22 minutes.

    \item Through our findings, we uncover key vulnerabilities in current approaches and propose a security-aware system model, a false positive mitigation strategy, and practical recommendations based on our five research questions to guide secure deployment.
    
    \item We publicly release our systems, attacks, and dataset to strengthen the collaboration on the topic: 
    \url{https://github.com/spritz-group/WAINE}.
\end{itemize}

\begin{table}[t]
\centering

\vspace{-1em}
\caption{Comparison of our systems' contributions with the \ac{sota}.
}
\label{tab:comparison}
\scriptsize

\begin{tabular}{lccccc}
\hline
  \textbf{Related Works} &
  \textbf{Auth.$^1$} &
  \textbf{Ident.$^2$} &
  \textbf{\begin{tabular}{@{}c@{}}System\\Model$^3$\end{tabular}} &
  \textbf{\begin{tabular}{@{}c@{}}Threat\\Model$^4$\end{tabular}} &
  \textbf{\begin{tabular}{@{}c@{}}Security\\Study$^5$\end{tabular}} \\
\hline

\cite{
abu2020livedi,
ahmadi2021optimising
} 
& \fullcirc & & & & \\
\multirow{1}{3.5cm}{\cite{
abdennour2021driver,
azadani2020performance,
azadani2021driver,
girma2019driver,
shin2025driver,
kwak2016know
}} %
& & \multirow{1}{*}{\fullcirc} & & & \\
\cite{
kang2019automobile,
ravi2022driver
}
& \fullcirc & \fullcirc & & & \\
\cite{TSENG2023105571} & \fullcirc & & \halfcirc & & \\
\cite{del2014real, el2019improving} & & \fullcirc & \halfcirc & & \\

\textbf{Ours} & \fullcirc  & \fullcirc  & \fullcirc  & \fullcirc & \fullcirc \\

\hline
\multicolumn{5}{l}{$^1$ Authentication. \; $^2$ Identification.} \\ 
\multicolumn{5}{l}{$^3$ Whether the paper considers issues of practical implementation.} \\ 
\multicolumn{5}{l}{$^4$ Whether the paper considers and discusses external threats' impact.} \\ 
\multicolumn{5}{l}{$^5$ Whether the paper considers possible attacks and needed mitigations. %
} \\ 
\end{tabular}
\vspace{-2em}
\end{table}

\section{Related Works}
\label{sec:relatedworks}

Behavior data is widely studied to identify the driver behind the wheel, which can support insurance and fleet monitoring, enable personalized driving experiences, and strengthen authentication for theft prevention.
For instance, it can adjust vehicle parameters based on user preferences or support driver-based insurance calculating risk factors~\cite{cura2020driver, nyt_behavior}.
While some works focus on extracting driving styles (e.g., aggressive, mild, or gentle)~\cite{cura2020driver}, the main goal remains driver identification and legitimacy.

\emph{Data.}
Most approaches extract \ac{can} bus data via the \ac{obd}-II port, providing sensor and actuator readings to characterize vehicle state.
While many studies do not share their datasets, some use publicly available ones, with the OCSLab dataset being the most widely used~\cite{martinelli2020human_ocslab, kwak2016know}, containing 54 sensor readings from ten drivers.
Other datasets include stability~\cite{ahmadi2021optimising} and a few incorporate physiological data~\cite{schneegass2013data}, though its real-world applicability is limited.
New datasets have also been released recently, but they are not widely adopted yet~\cite{wei2025pdb}.
Our work uses the OCSLab dataset to ensure comparability with existing literature.

\emph{Models.}
Most works on driver identification use classic \ac{ml} or advanced \ac{dl} algorithms.
Erzin et al.~\cite{erzin2006multimodal} first combined multimodal driving signals (e.g., pedal pressure, speed, steering) for pre-drive authentication.
In~\cite{ahmadi2021optimising}, a \ac{ml} model authenticates drivers and extracts features like gender.
Other \ac{ml} models, such as \ac{knn}~\cite{lin2018driver, ezzini2018behind, rahim2019zero} and \ac{rf}~\cite{ahmadi2021optimising, ezzini2018behind, rahim2019zero} have had varied success.
\ac{dl} models, especially \ac{lstm}~\cite{azadani2020performance, ravi2022driver, abu2020livedi, el2019improving} and \ac{rnn}~\cite{el2019improving, zhang2019deep, girma2019driver}, show better performance. Other approaches
, Autoencoder~\cite{chen2019driver}, ResNet~\cite{shin2025driver}, and \ac{gan}~\cite{park2019car}, have also been explored.

\emph{Features.}
The number of features used for driver identification has also been analyzed in the literature.
Marchegiani et al.~\cite{marchegiani2018long} tested single features (e.g., acceleration or brake) with \ac{svm}.
Rahim et al.~\cite{rahim2019zero} used only three \ac{gps}-based features (orientation change, stable speed, total acceleration) to achieve 90\% accuracy.
\ac{dl} models, better suited for handling many features, have been used with 51 features by Chen et al.~\cite{chen2019driver} in an Autoencoder and 40 features by Ravi et al.~\cite{ravi2022driver} in an \ac{lstm}.
Our study considers all features that can be extracted via the \ac{obd}-II interface, and we are the first to analyze the most relevant ones using \ac{xai} techniques. 

\emph{Implementation.}
Real-world implementation is a critical factor for these systems, yet many works fail to specify which components handle driver authentication.
Simple \ac{ml} algorithms can run on vehicles, but complex \ac{dl} networks typically need powerful servers, especially for training.
El Mekki et al.~\cite{el2019improving} demonstrated an authenticator using Automotive Grade Linux~\cite{automotiveGradeLinux} for UI personalization, but no detailed implementation of a physical anti-theft device exists.
Due to the high cost of \ac{ml} and \ac{dl} tasks, some works offload these to cloud servers. Ezzini et al.~\cite{ezzini2018behind} proposed remote servers to process \ac{can} bus data and authenticate drivers, while others~\cite{el2019improving, azadani2020performance, gahr2018driver, kang2019automobile} involve remote servers in authentication.
However, these studies overlook the need to secure vehicle-server communication and ignore privacy risks posed by third-party data sharing.
Our work explores the implementation of a lightweight behavior-based drive authentication for local use.  

\emph{Adversarial Attacks.}
There is a growing body of literature on adversarial attacks targeting \ac{ai}-enabled authentication systems.
Tan et al.~\cite{8852414} classified adversarial sample generation strategies into imitation-, surrogate-, and statistics-based approaches, while Marrone et al.~\cite{8987399} demonstrated that adversarial perturbations can successfully bypass both liveness detection and authentication in \ac{fas}.
However, we are the first to consider the effect of adversarial attacks against behavior-based driver authentication systems and their repercussions.

\section{System Model}
\label{sec:systemmodel}

We first discuss the physical implementation of the authenticator device\footnote{From here on, we will refer to ``authentication system'' and ``identification system'' interchangeably. From Section~\ref{sec:evaluation} we differentiate the tasks and treat them separately.} in Section~\ref{subsec:implementation}.
We then explain the data collection technique in Section~\ref{subsec:datacoll}.
Finally, we define training and testing procedures in Section~\ref{subsec:authentication}.

\subsection{Physical Implementation}
\label{subsec:implementation}

A common approach in the literature is to collect \ac{can} bus data from a device connected to the \ac{obd}-II port~\cite{park2019car, abu2020livedi, kang2019automobile}, typically by sending \ac{obd} requests and decoding the responses.
However, this setting is problematic for real-world driver authentication.
Since the authenticator is connected to an easily accessible port, an attacker inside the vehicle could simply remove it, much like unplugging a USB device, especially given the lack of anti-tampering mechanisms in current proposals. 
Additionally, existing models and the basic connection to the \ac{obd}-II port assume that messages exchanged between the gateway and the authenticator are unencrypted and lack integrity checks, leaving the system vulnerable to message tampering and injection attacks.

\begin{figure}[tbh]
    \vspace{-1em}
    \centering
    \includegraphics[width=.5\linewidth]{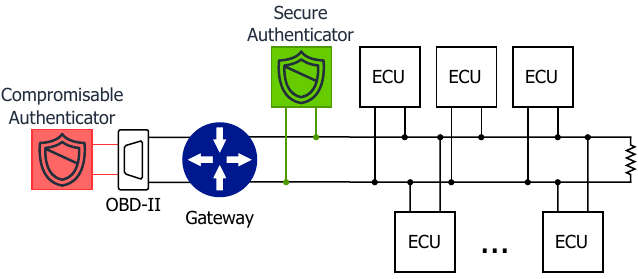}
    \caption{Examples of compromisable and secure implementations of the authenticator.}
    \label{fig:system}
    \vspace{-1em}
\end{figure}

For these reasons, we argue that \textit{the most secure and efficient (by default) implementation of the authentication system is directly on the \ac{can} bus, with the authenticator integrated as an \ac{ecu}}, as illustrated in Figure~\ref{fig:system}.
Placing the authenticator on the \ac{can} bus allows it to observe all network traffic and access sensor data from any \ac{ecu}.
This strategic location also makes tampering significantly more difficult and risky, as interfering with the \ac{can} bus could disrupt vehicle functionality and endanger safety. 
Furthermore, by embedding the authenticator \ac{ecu} in hard-to-reach areas, we reduce the risk of compromise, since accessing it would require considerable time and substantial expertise. Additionally, anti-tampering mechanisms can be implemented to alert in case of an authenticator disconnection~\cite{chan2014cyber}.

\textbf{RQ1 Takeaway} -- 
Behavior-based driver authentication systems connected to the \ac{obd}-II port are an easy target for tampering and disconnection. Deploying them as additional \acp{ecu} on the \ac{can} bus is recommended to reduce the risks of tampering and bypassing.

\subsection{Data Collection}
\label{subsec:datacoll}

While the \ac{can} bus is a well-known industry standard, each \ac{ecu} and vehicle \ac{oem} encodes data differently.
Since manufacturers do not publicly release decoding procedures, researchers must manually reverse engineer \ac{can} messages to build DBC files that define how to interpret the data~\cite{dbc-database}. This decoding process can be complex and time-consuming for researchers or third parties attempting to interact with the vehicle~\cite{choi2021enhanced}. In contrast, the \ac{oem} inherently possesses full access to the decoding information required to retrieve and interpret all relevant data.

In our system model implementation, we assume the authenticator has prior knowledge of how to decode the \ac{can} bus messages and is aware of which message IDs correspond to which \ac{ecu}.
This is a reasonable assumption, as a real-world implementation would likely be developed in collaboration with the vehicle manufacturer and the \ac{ecu} \ac{oem}.
Thus, data are collected directly from the \ac{can} bus, where the authenticator filters messages by ID and decodes them to extract relevant features for training and testing. To support continuous in-drive authentication, features are periodically sampled, and for \ac{dl} models exploiting temporal dependencies, the data are segmented into time windows and organized into batches.

In our implementation, we extract 46 features per second.
For our \ac{dl} model, we aggregate data into 16-second time windows with an 8-second step.
Time windows are then aggregated in batches of 4, resulting in an identification or authentication prediction every 40 seconds.
In contrast, the classical \ac{ml} model generates a prediction every second using individual samples. We discuss the trade-offs between these architectures in Section~\ref{subsec:baseline}.

\subsection{Authentication}
\label{subsec:authentication}

Following a new vehicle purchase or the start of driver training, the authenticator enters a training mode, collecting data batches to train the model. 
To preserve user privacy, we argue that \emph{data collection and model training should be performed in the driver's car}.
As we show in Section~\ref{subsec:baseline}, our models can be trained efficiently on low-resource devices, eliminating the need for external computing infrastructure. Once training is complete, the model is stored onboard for real-time classification.

During the operational phase (testing), the authenticator monitors the driver's behavior and periodically produces an authentication response.
A positive result indicates a legitimate driver and requires no action. 
A negative result, however, suggests behavioral deviation and can trigger an alert to the vehicle owner. 
The manufacturer or owner can choose the action to take in the event of unauthorized driving, such as causing the car to stop gracefully or producing an alarm (e.g., via the horn or the built-in alarm system). 
However, since such actions can pose safety hazards, we argue that \emph{the best course of action is to notify the vehicle owner, who can track the vehicle remotely and contact the authorities}.

To further reduce false positive alerts, we propose delaying notifications to the vehicle owner until multiple consecutive instances of unauthorized behavior are detected, rather than triggering an alert after a single batch of suspicious data. 
The reasoning is that with high authentication accuracy, the likelihood of multiple consecutive false positives, where legitimate driving behavior is mistakenly flagged as unauthorized, is low. We present an analysis of this rate, based on our system’s accuracy, in Section~\ref{subsec:baseline}.

\section{Threat Model}
\label{sec:threatmodel}

Potential attackers can gain physical access to a vehicle's \ac{can} bus network by exploiting its lack of built-in security, a method already used in real-world scenarios to inject packets and bypass vehicle security for unauthorized use, not merely theft~\cite{headlights}.
Indeed, despite the existence of many CAN bus tailored authentication protocols \cite{lotto2024survey},  real-world automotive CAN bus implementations do not adopt them.
Our proposed implementation assumes the authenticator device is securely installed in less accessible locations, such as inside the transmission tunnel or within a sealed compartment behind the engine bay firewall, preventing direct tampering.
Within this context, the adversary deploys a covert malicious device directly into the vehicle's \ac{can} bus network in an easily reachable location, such as the external headlights~\cite{headlights}.
This specialized device, capable of executing the perturbation function and featuring sufficient memory and computational power, may be a Raspberry Pi 4B, a low-cost and widely available device.
Furthermore, it could include \ac{gps}-enabled functionality if the attacker needs to track the victim's location (only in a specific case of our SMARTCAN attack). 
This device reads frames from the \ac{can} bus, applies adversarial perturbations on each frame, and injects them into the bus.
Alternatively, an attacker can remotely compromise an exposed \ac{ecu}, establish a connection to the bus, and use it as a proxy to launch attacks.

We also assume the attacker knows, at least in part, the vehicle's DBC file and thus can encode and decode data at will.
Indeed, the reconstruction of the DBC file can be performed manually or with automatic methods~\cite{buscemi2023survey, choi2021enhanced}. 
Moreover, many reversed DBC files are available online with frequent additions and updates~\cite{dbc-database}.
The attacker derives the features used for authentication from the DBC file, which provides detailed information about the data exchanged in \ac{can} messages.
By analyzing these messages, the attacker can identify the specific signals and parameters likely used by the model, enabling them to approximate the model's authentication features.
Furthermore, many works in the literature consider the same set of features, which appear to be the most effective for this task~\cite{girma2019driver,ravi2022driver,zhang2019deep}, as we will discuss in Section~\ref{subsec:dataset}. 

It is worth noting that the attacker must exercise caution, as injecting data with all features may not be feasible.
Injection of data into the \ac{can} bus poses inherent safety risks, given the potential impact on vehicle operation and driver safety.
To address this concern, we consciously selected specific features to enable controlled data injection while minimizing safety hazards.
Of the 46 features in our dataset, we identify 22 that can be manipulated without affecting the vehicle's behavior (e.g., engine coolant temperature, intake air pressure), while altering the remaining 24 could pose a risk to the driver (e.g., current gear, throttle position signal).
Table~\ref{tab:safety_features} shows examples of modifiable, borderline, and non-modifiable features.
We marked features as borderline whose modification may affect the vehicle's physical state; these will be assessed in future work.

\begin{table}[tbh]
\vspace{-1em}
\centering
\footnotesize
\def\arraystretch{1}
\caption{Distribution of features into safety classes.}
\label{tab:safety_features}

\resizebox{\columnwidth}!{
\begin{tabular}{l|c|l}
\hline
\textbf{Importance} & \textbf{\#} & \textbf{Examples} \\ \hline
\rowcolor[HTML]{D9EAD3} 
Modifiable     & 22 & \begin{tabular}[c]{@{}l@{}}Engine coolant temperature; Intake air pressure; Calculated road gradient\end{tabular} \\ \hline
\rowcolor[HTML]{FFF2CC} 
Borderline     & 15 & \begin{tabular}[c]{@{}l@{}}Engine speed; Acceleration speed - Lateral; Torque converter speed\end{tabular}        \\ \hline
\rowcolor[HTML]{F4CCCC} 
Non-modifiable & 9  & \begin{tabular}[c]{@{}l@{}}Throttle position signal; Current gear; Master cylinder pressure
\end{tabular}
\\ \hline
\end{tabular}
} %
\end{table}

\vspace{-1em}

\section{Methodology}
\label{subsec:gan-can}

We provide details on the dataset used and its processing in Section~\ref{subsec:dataset}, while we detail our models in Section~\ref{subsec:authenticator}.
Then, we present our two attacks, \textbf{SMARTCAN} and \textbf{GANCAN}, in Section~\ref{subsec:attacks}.

\subsection{Dataset}
\label{subsec:dataset}

To maintain consistency with prior research, we use the widely adopted OCSLab dataset~\cite{martinelli2020human_ocslab, kwak2016know}.
This is the most widely used dataset in the literature discussed in Section~\ref{sec:relatedworks} and remains a benchmark for behavior-based driver authentication systems due to its comprehensive and structured data.
It includes approximately 26 hours of driving data
from 10 different drivers, from which 54 distinct features have been extracted.
These features are extracted from messages on the \ac{can} bus and are highly correlated with driving behavior, e.g., acceleration, speed, brake switch indication, and fuel consumption.
Eight constant/zero-valued features were removed, leaving 46 features for authentication.
We applied a windowing technique to the time series data, segmenting it into overlapping windows with a step size of half the window length. 
We chose a 16-second window, shorter than typical values in related work~\cite{abdennour2021driver, abu2020livedi}, to enable more frequent classification and finer-grained analysis of driving behavior.

Preprocessing included normalization to mitigate scale biases and improve data quality for authentication.
To address class imbalance, we undersampled the majority classes, resulting in a balanced dataset of 74,200 samples.
This enhances the model's ability to learn from both majority and minority classes.

\subsection{Implemented Models}
\label{subsec:authenticator}

Most related works do not share their code openly, and only a fraction include enough details in the manuscript to accurately replicate the results.
As such, we focused on the ones yielding the best-claimed results, i.e., \cite{girma2019driver,ravi2022driver,zhang2019deep}.
For our original implementations, we investigated different architectures, such as Decision Trees, \acp{svm}, XGBoost, and Hidden Markov models, but focused our evaluation on our best-performing models. 
All the implementation code is available in our GitHub repository.

\emph{Our \ac{rnn}.}
\label{subsub:ourdl}
We designed a simple, computationally lightweight \ac{rnn}-based driver authenticator. The behavior-based model uses a single bidirectional \ac{gru} layer with 220 units to capture temporal dependencies, followed by a dropout (rate 0.1) to mitigate overfitting. A linear layer with 440 neurons and a sigmoid activation produces an output vector sized to the number of drivers. Inputs are sequential driving data arranged as a 3D tensor with shape (batch size, window size, feature count). This architecture prioritizes temporal modeling while remaining practical and resource-efficient for deployment.
We adopt the cross-entropy loss for multi-class classification and train the model using the Adam optimizer with a learning rate of $1 \times 10^{-3}$ for 10 epochs.

\emph{Our \ac{rf}.}
\label{subsub:ourml}
This type of model has been used in several other works on identification and authentication and in automotive-related tasks~\cite{marchiori2023your, lee2019automotive}.
After hyperparameter tuning, our model consists of 100 estimators, using Gini impurity as the criterion.
Being a \ac{rf} model, it is computationally lightweight for both training and inference, making it well-suited to our constrained scenario, and it produces predictions on each collected sample rather than requiring a full window. Consequently, the system provides continuous authentication at one prediction per second of driving.

\subsection{Attacks}\label{subsec:attacks}
Based on the attacker's knowledge, we design two different methodologies.
SMARTCAN describes the situation where the attacker has access to data originating from a legitimate user's driving.
GANCAN, instead, details the attacker's strategy for accessing the model's output.
While not adhering to the traditional \ac{gan} training framework, our generator models (when used) are optimized to bypass a fixed discriminator (i.e., the authentication system) rather than incrementally training both components.
Thus, the deployed generative networks detailed in this section leverage the discriminator feedback (i.e., the authentication system output) for optimization and loss computation.

\subsubsection{SMARTCAN}
We assume the attacker can access the legitimate driving behavior but does not know the authenticator model implementation.
One way to access it involves eavesdropping on an insecure channel where data is transmitted to third parties, or through possible leaks from companies collecting behavioral data~\cite{cheng2017enterprise}.
This might be the case for insurance companies using \ac{ubi} or for rental companies monitoring and storing their clients' driving behavior~\cite{nyt_behavior}.
However, if the attacker cannot access the data, they can physically deploy a malicious device to be connected to the \ac{can} bus of the target vehicle.
The device sniffs the traffic and extracts the driving behavior, which is then sent to the attacker.
Once enough data has been collected, the attacker can launch the attack using the same device.

Due to the feature safety reasons detailed in Section~\ref{sec:threatmodel}, the attacker cannot simply replay the legitimate traffic they have access to.
Thus, we propose a \textit{smart}-replay attack in which the attacker can replay legitimate traffic using only the features labeled as \emph{modifiable}.
Instead, the data for all other features (i.e., \emph{borderline} and \emph{non-modifiable}) are taken from the attacker's driving behavior during the car theft.
In this way, the attacker can drive the car without any interference with their driving safety and still be authenticated by the anti-theft system.

\subsubsection{GANCAN}
In this attack, we assume the attacker can access the authenticator model response but does not know the legitimate driving behavior. 
An attacker may obtain the trained model by dumping \acp{ecu} memory remotely~\cite{cardump_reddit} or physically~\cite{dump_rom_guide} (e.g., colluded workshop personnel dumping models to smuggle them). If trained on the cloud, it may be leaked when transmitted back to the vehicle~\cite{gong2020model}
or, if used as a system-as-a-service, when asked to evaluate a vehicle’s state.
Other options include obtaining data on a similar car by colluding with a car-sharing platform and using it to train a model that mimics the vehicle’s authenticator~\cite{chen2023adaptive}.

Instead, if the attacker has no access to a copy of the model, they can use the malicious device connected to the victim's vehicle to access the model while the vehicle is idle and parked. 
Indeed, while the car is stationary, the authenticator remains operational but does not alert the owner in case of unauthorized access.
This happens because \ac{can} traffic would inherently be different for a parked vehicle with respect to a vehicle moving on the road.
It may be possible in cases where the authenticator is developed by a third party and integrated into the car by the manufacturer, utilizing a separate \ac{ecu} to generate actions in response to alerts. In such a case, the authenticator must avoid taking actions when data is invalid, such as when the vehicle is parked. 
Despite being more time-consuming, the attacker can perform the attack in just one stage.
Indeed, as we will see in the evaluation section (Section~\ref{sec:evaluation}), the authentication results are broadcast at fixed intervals.

The attacker can preemptively use the legitimate model to train the generator of a \ac{gan}.
Indeed, it is possible to use the authenticator model as an oracle and train a neural network on its feedback, obtaining a model that can craft legitimate-looking fake packets from noise.
Our attack aims to create authentic features that resemble genuine drivers without compromising safety by overwriting only \textit{modifiable} features.
This allows the \ac{gan} to generate complete batches of data, of which the validity is required only for the \textit{modifiable} features that are injected in the \ac{can} bus.
\textit{Non-modifiable} and \textit{borderline} features are instead naturally extracted from the attacker's driving behavior.

\begin{figure}[tbh]
    \centering
    \includegraphics[width=.5\linewidth]{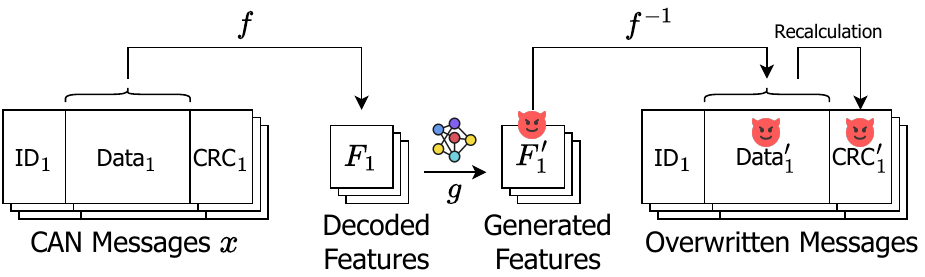}
    \caption{Malicious \ac{can} message generation.}
    \label{fig:validation}
    \vspace{-1em}
\end{figure}

GANCAN packet generation and injection comply with \ac{can} bus format specification~\cite{canbus_standard}.
Our generator model operates on the feature space of the \ac{can} bus packet data field to avoid overwriting data that does not need to be modified in our attack (e.g., arbitration IDs).
In particular, let us define as $\bm{x}$ the frame/batch of frames used as input to the model.
Denoting as $f(\bm{x})$ the function with image $\mathcal{F}$ mapping $\bm{x}$ to its feature space, our generation process $g$ is such that $g(f(\bm{x})) \to \mathcal{F}$.
In other words, our generation process ensures that the data we generate complies with the data expected from that specific feature.
In the context of the \ac{can} bus, the function $f$ represents the decoding procedure that converts the raw bits from the message data field to the numeric values of a feature used for authentication.
Since \ac{can} message encoding is linear and detailed in each vehicle DBC file, it is represented as $f^{-1}$.
The original data field of the packet (i.e., the one generated by the attacker's driving) is overwritten with the data generated via our model.
Then the fields that depend on the frame content (e.g., the CRC field) are updated.
This ensures that our generated packets are \textit{(i)} compliant with the \ac{can} bus standard and \textit{(ii)} in a reasonable range for the specific generated feature.
An overview of the generation process is shown in Figure~\ref{fig:validation}.

Given the vast search space and the potential challenges of convergence and local minima, we employ \ac{rl} to optimize the learning procedure of our generator.
Doing so enables the generator's feedforward training algorithm to explore the search space more efficiently and adapt its strategy based on feedback and rewards during learning.
The \ac{rl} parameters include the maximum episode length, the number of episodes, the learning rate $\alpha$, and the discount factor $\gamma$.
These parameters govern how the generator's latent input will be updated using \ac{rl}.
The training loop consists of episodes, where each episode starts with initializing the latent input and the episode reward.
The generator generates a sample within each episode based on the latent input.
This generated sample is then evaluated by the surrogate model (i.e., the authentication system).
The surrogate model's output is used to make predictions, and random target labels are generated for comparison. The RL loop uses random targets to first locate the manifold of realistic driving data. Only once the generator has found that region does the cross-entropy step steer it toward the specific target driver, avoiding the much harder problem of optimizing directly from pure noise.
The reward is calculated as the mean accuracy of predictions that match the targets. The latent input is updated using \ac{rl} by computing the temporal difference error (i.e., $td_{error}$) as the difference between the reward and the cumulative episode reward.
The latent input is then updated by adding a scaled noise term to introduce randomness and exploration.
The scaling factor is determined by $\alpha$, $td_{error}$, and the $\gamma$ factor raised to the power of the current step:
 $   \alpha \cdot td_{error} \cdot \gamma^{\,step} \cdot latent\_input$.
This update process helps the generator adjust its latent input based on the reward signal and explore different regions of the latent space.
After the episode, the generator is updated using the final latent input.
The surrogate model evaluates the generator's output, and a target label tensor is created for the loss calculation.
The generator's loss is computed using the cross-entropy loss function, and backward propagation is performed to update the generator's parameters.
We use the Leaky ReLU activation function~\cite{xu2020reluplex} to prevent the dying ReLU problem and to improve gradient flow, which in turn helps stabilize the training of our generator.
The generator consists of 5 feed-forward layers with 128, 256, 512, 64, and 46 units, respectively.
Figure \ref{fig:gancatrain} in the Appendix shows the training steps.

\section{Evaluation}
\label{sec:evaluation}

We now evaluate our driver authentication and identification systems,
encompassing all scenarios detailed in the previous sections.
As metrics, we use accuracy and F1 score to evaluate the models, and \ac{asr} to evaluate attacks, where we define \ac{asr} as the ratio of malicious batches that fool the authenticator to the total number of malicious batches sent.

We include results for the identification and authentication tasks for each evaluation in this section.
While the data and features utilized for both tasks remain consistent, fundamental conceptual distinctions set them apart:
\begin{itemize}
    \item \emph{Identification}: determining which specific driver is currently operating the vehicle.
    Thus, this task assigns each driver a unique identity within a set of known drivers.
    As a \ac{ml} classification problem, this translates to a multiclass classification task where each label constitutes a driver.
    \item \emph{Authentication}: verifying whether the current driver is legitimate.
    Thus, this task confirms the driver(s)' legitimacy based on behavior characteristics.
    As a \ac{ml} classification problem, this translates to a binary classification task.
\end{itemize}

\noindent We set the baseline for the models and attacks by testing \ac{sota} and our systems under no attack (Section~\ref{subsec:baseline}) and in specific attack scenarios (Section~\ref{subsec:attacksevaluation}).

\subsection{Baseline Models Evaluation}
\label{subsec:baseline}
We now evaluate the baseline performance of our behavior-based driver identification and authentication systems in the absence of attacks.
We divide our dataset into two subsets: training (80\%) and test (20\%).
We train and test all the models considered (i.e., our implementations and those from the literature) on the same dataset.
We employed the same time-window-based preprocessing across all models, except for our \ac{rf} model, which processes individual samples, as detailed in Section~\ref{subsec:datacoll}.
We summarized the results of our baseline evaluation in Table~\ref{tab:models_comparison}.

\emph{Results.}
Our models outclass the \ac{sota} on most tasks, regardless of architecture.
The \ac{rf} results are notable as they capture dataset patterns without relying on sample-to-sample causality.
Thus, feature interactions are as crucial as temporal dependencies between samples.
Therefore, while understanding parameter evolution can still aid driver identification (as seen by the \ac{rnn} model performance), our \ac{rf}, excels due to feature correlations unique to each driver (e.g., acceleration during turns) which is consistent with prior evidence that tree-based models remain highly competitive on medium-sized tabular data~\cite{grinsztajn2022tree}.
We also observe a general improvement from identification to authentication, consistent with results for related tasks in the literature~\cite{marchiori2023your}.
For authentication, our dataset was rebalanced to maintain consistency in data processing (i.e., the number of samples for the positive and negative labels was equal).

\emph{Combinatorial Accuracy.}
As noted in Section~\ref{sec:systemmodel}, the authenticator’s behavior depends on several system parameters.
While our systems demonstrate high accuracy in evaluation, handling misclassifications is essential for real-world deployment.
Alerting the owner on every unauthorized batch detection would generate excessive false alarms.
To reduce false alarms, we require several consecutive unauthorized batches before issuing an alert.
We introduce \emph{combinatorial accuracy}, which quantifies the number of consecutive unauthorized batches required to trigger an alert.
This concept quantifies the likelihood of successfully detecting unlawful drivers while ensuring system robustness.
The formula for combinatorial accuracy reflects a geometric distribution as follows.

\begin{equation*}
    combinatorial\_acc = \left( 1 - \left( 1 - accuracy \right)^{batches} \right).
\end{equation*}

By doing so, we ensure that a higher accuracy value is used for each notification.
An analysis of the false alarm probability as a function of the number of consecutive unauthorized batches is shown in Table~\ref{tab:combinatorial}.
As can be seen, including a second unauthorized batch of data, the probability of false alarm drops to zero for both our \ac{rnn} and \ac{rf} models.
Hence, we select two batches of unauthorized data as parameters for the notification delay, since they strike a fair trade-off between false alarm rate and collection time.
It is also worth noting that, in the case of our \ac{rf} model, we replicate the same batch-level evaluation used for the \ac{rnn}, grouping test samples into non-overlapping batches of four and applying a majority vote over the individual predictions to produce a single batch-level decision.

\begin{table}[tbh]
\centering
\scriptsize
\def\arraystretch{1}

\begin{minipage}[t]{0.49\textwidth}
\centering
\vspace{-1em}
\caption{Comparison of our model's performance with the \ac{sota}.}
\resizebox{\textwidth}{!}{ %
\label{tab:models_comparison}
\begin{tabular}{l|cc|cc}
\hline
\multirow{2}{1cm}{\textbf{Model}} &
\multicolumn{2}{c|}{\textbf{Identification}} &
\multicolumn{2}{c}{\textbf{Authentication}} \\ \cline{2-5}
& \textbf{Accuracy} & \textbf{F1} & \textbf{Accuracy} & \textbf{F1} \\
\hline
Girma et al.~\cite{girma2019driver} & 0.987 & 0.988 & \textbf{0.999} & \textbf{0.999} \\
Ravi et al.~\cite{ravi2022driver} & 0.995 & 0.996 & 0.999 & 0.999 \\
Zhang et al.~\cite{zhang2019deep} & 0.848 & 0.847 & 0.989 & 0.989 \\
\textbf{\ac{rnn} (ours)} & \textbf{0.998} & \textbf{0.999} & \textbf{0.999} & \textbf{0.999} \\
\textbf{\ac{rf} (ours)} & \textbf{0.998} & \textbf{0.998} & \textbf{0.999} & \textbf{0.999} \\
\hline
\end{tabular}
} %
\vspace{-1em}
\end{minipage}
\hfill
\begin{minipage}[t]{0.49\textwidth}
\centering
\vspace{-1em}
\caption{False alarm mean probability for identification when delaying notification by 
\#X
consecutive unauthorized batches.}
\label{tab:combinatorial}
\resizebox{\textwidth}{!}{ %
\begin{tabular}{l|ccc}
\hline
\multirow{2}{1cm}{\textbf{Model}} &
\multicolumn{3}{c}{\textbf{False Alarm Probability}} \\ \cline{2-4}
& \textbf{Batch \#1} & \textbf{Batch \#2} & \textbf{Batch \#3} \\
\hline
Girma et al.~\cite{girma2019driver} & 1.996e-02 & 0.00000 & 0.00000 \\
Ravi et al.~\cite{ravi2022driver} & 8.869e-03 & 0.00000 & 0.00000 \\
Zhang et al.~\cite{zhang2019deep} & 2.018e-01 & 5.405e-02 & 2.041e-02 \\
\textbf{\ac{rnn} (ours)} & \textbf{4.435e-03} & \textbf{0.00000} & \textbf{0.00000} \\
\textbf{\ac{rf} (ours)} & \textbf{6.652e-03} & \textbf{0.00000} & \textbf{0.00000} \\
\hline
\end{tabular}
} %
\vspace{-1em}
\end{minipage}
\end{table}

\emph{Local Training Feasibility.}
We evaluate the training and inference feasibility of our models on a constrained environment to simulate running on a vehicle's \ac{ecu}.
We emulate it using a Raspberry Pi 4B~\cite{rpi} 
without any optimization in the hardware or on the code that may speed up the computations (e.g., rewriting in a compiled language). 
It took an average of 103 seconds to train the \ac{rf} identification models, while about 67 minutes to train our \ac{rnn} identification model. 
For inference, \ac{rf} evaluates one sample in 0.029 seconds, while \ac{rnn} processes a batch in 0.010 seconds.
Because \ac{rf} runs per sample (every 1 second) and RNN runs per 40-second batch, we are well within limits for real-time use.
In practice, the reported training time should be interpreted as a cumulative cost rather than a single uninterrupted session, as the RNN can be trained incrementally across multiple trips and during vehicle downtime.

\subsection{Attacks Evaluation}
\label{subsec:attacksevaluation}

To assess the effectiveness of our attacks, we separately evaluate the cases where the attackers have access to the data (SMARTCAN) or to the model (GANCAN). 
We also include an analysis of our attacks across different numbers of modifiable features, along with a final summary showing the effectiveness in terms of success rate and time required to steal the vehicle (Table~\ref{tab:summary}).

\subsubsection{SMARTCAN Evaluation}
\label{subsub:data}

To evaluate this attack, we consider each driver both as an attacker and as a victim.
We overwrite the \emph{modifiable} features for each attacker data batch with those from the victim driver's batches (target).
We then feed these combined data batches to all our models.
The purpose of this evaluation is to determine if the models can classify the combined data as the same class as the target driver (i.e., \emph{modifiable} features from the target driver data, \emph{borderline} and \emph{non-modifiable} features from the attacker's driving behavior data).

As shown in Table~\ref{tab:gb1}, our attack achieves an \ac{asr} high enough to disrupt system functionality.
Moreover, in this task, we observe a correlation between the \ac{asr} and the models' baseline accuracies and F1 scores (i.e., the higher the system's accuracy, the more susceptible it is to the attack).
Instead, our \ac{rf} model appears to be among the most resilient models.
This behavior can be explained by its reliance on feature interaction rather than temporal dependency.
Indeed, \ac{rnn} models rely on the causality of the samples and, as such, might focus on the time evolution of specific features that might be maliciously injected.
Our \ac{rf} model instead used only the interaction between features in each data sample to generate its prediction, and thus ``mixed'' samples (with \textit{modifiable} features from one driver and \textit{borderline} and \textit{non-modifiable} features from another) can be detected more easily.

\begin{figure}[tbh]
\vspace{-1em}
\centering
\begin{minipage}[]{.45\textwidth}
    \centering
    \includegraphics[width=.9\linewidth]{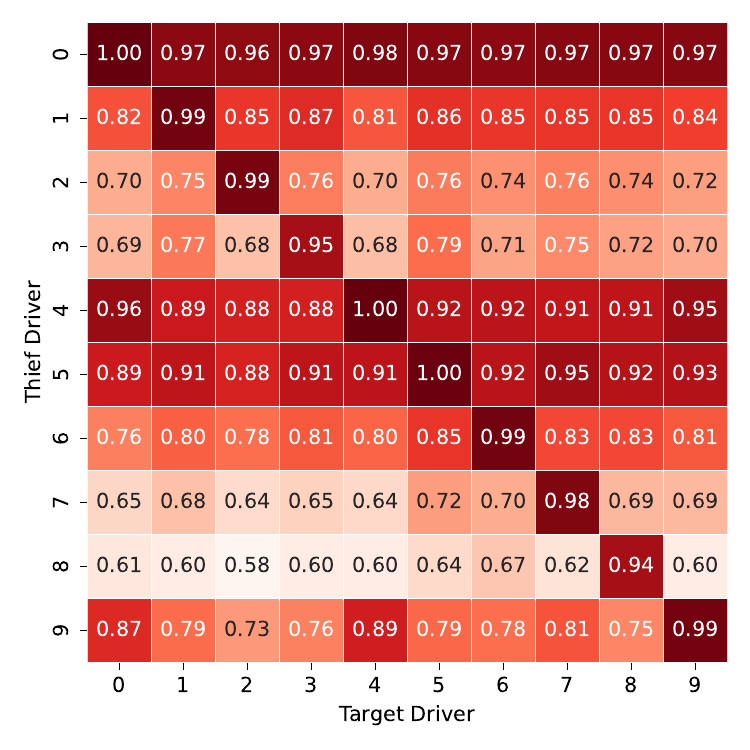}
    \caption{SMARTCAN \ac{asr} for each target and attacker drivers based on \ac{rf}.
    }
    \label{fig:gb1_eval}
\end{minipage}
\hspace{1em}
\begin{minipage}[]{.45\textwidth}
    \centering
    \includegraphics[width=\linewidth]{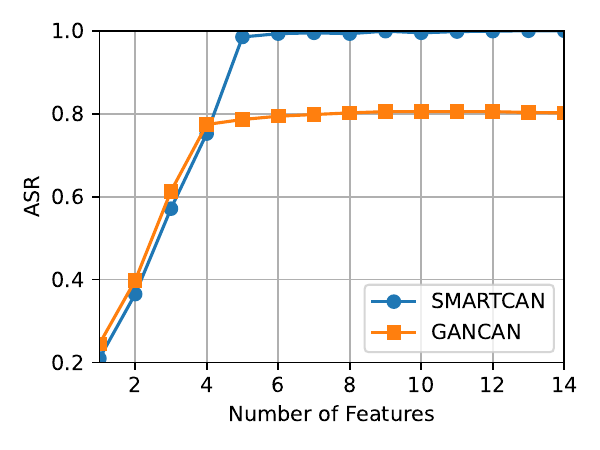}
    \caption{\ac{asr} of our attacks when considering different numbers of injectable features.}
    \label{fig:refined}
\end{minipage}

\vspace{-1em}
\end{figure}

Nevertheless, our attack results show that identification is vulnerable regardless of whether \ac{rnn} or \ac{rf} models are used.
We show the \ac{asr} for each combination of thief and target drivers for our \ac{ml} model in Figure~\ref{fig:gb1_eval}.
It can be noticed that some drivers are more vulnerable than others (e.g., drivers 0, 4, and 5).
This can be due to several factors, such as unusual driving patterns in the training data or biases in dataset creation (e.g., different routes and collection times).
Conversely, some drivers (e.g., driver 8) appear more resilient to the attack.

In the case of authentication, our attack shows perfect \ac{asr} (i.e., 1) against the \ac{rnn} models.
This highlights the heightened vulnerability of these models to these adversarial attacks.
Instead, the \ac{asr} on our \ac{rf} model reaches 0.739.
This behavior supports our hypothesis about the importance of feature interactions for the \ac{rf} model, as further evidenced by the decrease in the number of labels in the dataset.

\paragraph{Getting knowledge of the legitimate data.}
If the attacker needs to obtain data samples through the same device, it is crucial to determine the number of batches required for the smart-replay attack to be effective. 
To evaluate it, we consider different percentages of the original test set (from 10\% to 90\% in increments of 10\%) and repeat the evaluation.
The results show that the test set size does not significantly affect the attack, as the \ac{asr} drops by at most 0.002 and the standard deviation of the obtained \acp{asr} is at most 0.001 for the \ac{dl} models.
Therefore, the attacker can stop the first part of the attack after collecting just one batch of legitimate data and use it for all subsequent SMARTCAN attacks.
In the case of our \ac{rf}, we noticed that randomly selecting a sample from the test set for our attack is not the best strategy, as we obtained a mean \ac{asr} of 0.850 and a standard deviation of 0.230 (over multiple runs).
As such, the attack needs at least 10 data samples to be successful, which still constitutes an improvement over the 40 seconds required to collect a batch of data for a \ac{dl} model.

\begin{table}[tbh]
\vspace{-1em}
\centering
\def\arraystretch{1}
\scriptsize
\caption{\ac{asr} for \textbf{Ident}ification and \textbf{Auth}entication evaluated on all models.}
\label{tab:gb1}
\begin{tabular}{l|cc|cc}
\hline
  \multirow{2}{1cm}{\textbf{Model}} &
  \multicolumn{2}{c}{\textbf{SMARTCAN}} &
  \multicolumn{2}{c}{\textbf{GANCAN}} \\ \cline{2-5}
  & \textbf{Ident.} & \textbf{Auth.} & \textbf{Ident.} & \textbf{Auth.} \\
\hline

Girma et al.~\cite{girma2019driver} & 0.985 & 1.000 & 0.996 & 1.000 \\
Ravi et al.~\cite{ravi2022driver} & 0.992 & 1.000 & 0.963 & 1.000 \\
Zhang et al.~\cite{zhang2019deep} & 0.830 & 1.000 & 0.848 & 1.000 \\

\textbf{\ac{rnn} (ours)} & \textbf{0.998} & \textbf{1.000} & \textbf{0.998} & \textbf{0.903} \\
\textbf{\ac{rf} (ours)} & \textbf{0.817} & \textbf{0.739} & \textbf{0.590} & \textbf{0.818} \\
\hline
\end{tabular}
\vspace{-1em}
\end{table}

\subsubsection{GANCAN Evaluation}
\label{subsub:model}

To evaluate this attack, we employ a generator model optimized via \ac{rl} that uses the output of our systems as the surrogate model for each driver in the dataset.
We aim to generate data packages that could be classified as the authenticated driver.
For this evaluation, we will consider each driver in the dataset as the legitimate driver once in the case of identification.
For authentication, instead, we specifically aim to have the negative class labeled as positive.
We trained each generator to forge data tailored to each dataset driver.
Furthermore, we use the generated data only for the \emph{modifiable} features, while the others are extracted from the attacker's driving behavior.
We show the results of our evaluation in Table~\ref{tab:gb1}.
From these \acp{asr}, we notice a similar behavior to the case where the attacker knows the data (i.e., SMARTCAN).
Namely, models that perform best in the baseline evaluation are also more susceptible to attacks.
Furthermore, also in this case, authentication is the most vulnerable task, as \acp{asr} reach values up to 1.
Again, our \ac{rf} appears to be the most resilient against our attacks, but its accuracy still drops to a level too low to allow the system to function normally.
In GANCAN, the target authenticator serves as a fixed oracle, and the generator learns through reinforcement from its feedback. Consistent with prior observations on model sharpness and robustness~\cite{yu2018interpreting}, sharper and more locally consistent decision boundaries can make optimization easier by giving clearer boundary-crossing signals and faster convergence, whereas flatter, noisier, or piecewise-constant rewards, such as those from \ac{rf}, provide weaker guidance, slowing convergence and reducing attack effectiveness.

\paragraph{Getting knowledge of the model.}
If the attacker does not have offline access to the model, they can use the authenticator as an oracle for training the \ac{gan} generator.
Thus, the number of episodes the generator model needs to converge becomes tightly related to the time required to steal the car.
Indeed, for all tested \ac{dl} models, we can generate only one batch of data collected by the authentication system over 40 seconds per episode.
We test the number of episodes required for convergence for each target driver, yielding a mean of 30.
The time required to process each episode, even on the CPU, is negligible (around 0.1 seconds).
Therefore, the attacker will need, on average, 20 minutes (40 seconds for each of the 30 episodes) to train the generator and then inject malicious packets to fool the authenticator model.

The attack targeting our \ac{rf} model requires the same amount of time.
Indeed, while the architecture the attack is trying to fool is different, the surrogate model used to generate the attack is the same.
Even though authentication is performed every second, the generator still needs 20 minutes to converge.
After its deployment, each generated batch can be decomposed into individual data samples for injection.

To identify the most critical modifiable features for compromising the authentication model, we use 
\ac{shap}, a model-agnostic \ac{xai} technique~\cite{lundberg2017unified}.
14 of the top 20 features identified as important based on Shapley values on our \ac{rnn} model are modifiable features (see Appendix).
As such, we study the \ac{asr} of the two attack strategies when modifying fewer features.
Figure~\ref{fig:refined} shows the result of this study, wherein for each position $x$ we perform the attack with the $x$ most important \textit{modifiable} features.
The results show that the SMARTCAN attack can obtain an \ac{asr} of 0.993 with only 6 features.
The optimal number of features to modify for the GANCAN attack is 8, which achieves an \ac{asr} of 0.802.
Thus, for the attacker, injecting the 8 most important \textit{modifiable} features is the best tradeoff between \ac{asr} and driving ease.

\textbf{RQ2 Takeaway} -- All state-of-the-art models are vulnerable to our proposed attacks, rendering them not reliable for authentication or identification purposes.

\noindent\textbf{{Summary}}
\label{subsub:summary}
Across the paper, our threat model assumes the attacker’s primary goal is vehicle theft.
However, similar strategies could be employed to modify user behavior, such as committing insurance fraud.
We summarize strategies adversaries may employ based on their knowledge in Table~\ref{tab:summary}.
For attackers lacking system knowledge, the optimal attack depends on their physical access to the vehicle.
Getting access to the data is the stealthiest way to steal the vehicle if they can access it multiple times.
A \ac{gps} tracker is inserted into the malicious device to facilitate locating the vehicle in the second stage.
As shown, the attack is successful with a single batch of legitimate data, requiring only a few minutes.
When other security measures are in place, e.g., \acp{ads}~\cite{marchetti2017anomaly, lokman2019intrusion}, the attacker needs more data to mimic the driver.
Otherwise, an attacker can access the model while the victim is away and use GANCAN, requiring 22 minutes (setup and training of the generation model). 
When the attacker holds the victim's data, computation can be done offline, requiring 2 minutes to set up and launch the attack.

\begin{table}[tbh]
\vspace{-2em}
\def\arraystretch{1}
\centering
\caption{Summary of the attacks against our models. Values in parentheses are needed only if the attacker has no prior knowledge of the model or data.}
\label{tab:summary}
\resizebox{.7\columnwidth}!{
\begin{tabular}{c|c|c|c|c|c|c}\hline
\multirow{2}{*}{\textbf{Attack}} & \multirow{2}{*}{\textbf{Model}} & \multirow{2}{*}{\textbf{\ac{asr}*}} & \multicolumn{4}{c}{\textbf{Time to steal the vehicle}} \\ \cline{4-7}
 &
   & & 
  \textbf{Setup~\cite{headlights}} &
  \textbf{Data} &
  \textbf{Train} &
  \textbf{Total} \\ \hline

\multirow{4}{*}{SMARTCAN} & \multirow{2}{*}{\ac{rf}} & 0.817$_i$ & \multirow{4}{*}{2'} & \multirow{2}{*}{(10'')} & \multirow{4}{*}{-} & \multirow{2}{*}{\textbf{2' 10''}} \\ \cline{3-3}
& & 0.739$_a$ & & & & \\ \cline{2-3} \cline{5-5} \cline{7-7}
& \multirow{2}{*}{\ac{rnn}} & 0.998$_i$ & & \multirow{2}{*}{(40'')} & & \multirow{2}{*}{\textbf{2' 40''}} \\ \cline{3-3}
& & 1.000$_a$ & & & & \\ \hline

\multirow{4}{*}{GANCAN} & \multirow{2}{*}{\ac{rf}} & 0.590$_i$ & \multirow{4}{*}{2'} & \multirow{4}{*}{-} & \multirow{4}{*}{(20')**} & \multirow{4}{*}{\textbf{22'}} \\ \cline{3-3}
& & 0.818$_a$ & & & & \\ \cline{2-3}
& \multirow{2}{*}{\ac{rnn}} & 0.998$_i$ & & & & \\ \cline{3-3}
& & 0.903$_a$ & & & & \\ \hline

\multicolumn{7}{l}{*Subscripts $i$ and $a$ indicate, respectively, identification and authentication.} \\
\multicolumn{7}{l}{**If the attacker has previous access to the model, training can be done offline.}

\end{tabular}
} %
\vspace{-1em}
\end{table}

\noindent\textbf{{Discussion}} While the usage of behavior-based driver authentication systems has already been discussed in the previous sections, one might argue that physical keys will remain the primary access method, as system security depends on its weakest link. In addition, other second-factor authentication methods are available, for instance, by employing smartphones.
Yet, both are vulnerable to attacks~\cite{garcia2016lock, francillon2011relay, leu2022ghost}, and attackers can bypass them using various techniques~\cite{headlights}.
In contrast, driver authentication systems can detect, report, and stop intrusions even after initial access. They work best as a second-factor continuous authentication method, preserving key convenience while enhancing security. Beyond security, they enable applications such as \ac{ubi}~\cite{nyt_behavior, cura2020driver, meiring_review_2015}.

\noindent{\textbf{RQ3 Takeaway}} -- Behavior-based driver authentication systems can only serve as a second-factor authentication, and cannot represent the primary access method.

\section{Conclusions}
\label{sec:conclusions}

In this paper, we defined a realistic system and threat model for behavior-based driver authentication aligned with \ac{can}-bus constraints.
We then proposed two lightweight authentication approaches that achieve performance comparable to or better than more complex \ac{sota} solutions, with accuracies up to 100\%.
We also introduced \textbf{SMARTCAN} and \textbf{GANCAN}, the first attacks targeting behavior-based driver authentication under different assumptions about attacker knowledge, demonstrating that they can bypass authentication in under 22 minutes. 
While promising, validation on real vehicles would strengthen our findings, but it is hindered by the closed-source nature of automotive systems and restricted access to proprietary \ac{ecu} data formats.
A further limitation is that many formulations collapse multi-class identification into a binary task by grouping all non-target drivers into one negative class.
In reality, the negative class is far more diverse, but privacy and data availability constraints in academic datasets likely prevent prototypes from matching commercial performance.

\bibliographystyle{splncs04}
\bibliography{bibliography.bib}
\section*{Appendix}

\begin{figure}[tbh]
    \centering
    \begin{minipage}{0.48\linewidth}
        \centering
        \includegraphics[
        width=\linewidth,
        trim=2.5cm 0cm 2.5cm 0cm,
        clip
    ]{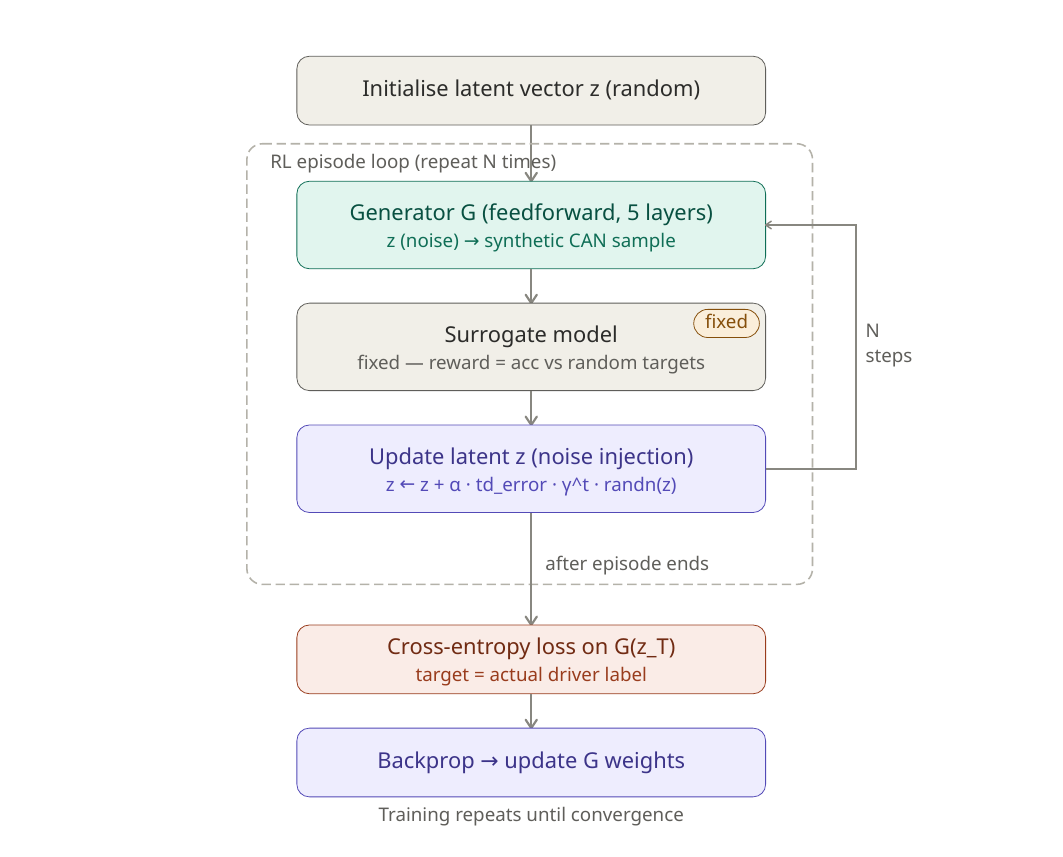}
        \caption{GANCAN training pipeline.}
        \label{fig:gancatrain}
    \end{minipage}
    \hfill
    \begin{minipage}{0.48\linewidth}
    
    {\scriptsize
    \begin{enumerate}
        \item \texttt{Long\_Term\_Fuel\_Trim\_Bank1} (modifiable)
        \item \texttt{Torque\_of\_friction} (modifiable)
        \item \texttt{Engine\_coolant\_temperature.1} (modifiable)
        \item \texttt{Intake\_air\_pressure} (modifiable)
        \item \texttt{Engine\_soacking\_time} (modifiable)
        \item \texttt{Maximum\_indicated\_engine\_torque} (modifiable)
        \item \texttt{Activation\_of\_Air\_compressor} (modifiable)
        \item \texttt{Accelerator\_Pedal\_value} (borderline)
        \item \texttt{Engine\_coolant\_temperature} (modifiable)
        \item \texttt{Master\_cylinder\_pressure} (non-modifiable)
        \item \texttt{Calculated\_LOAD\_value} (modifiable)
        \item \texttt{Indication\_of\_brake\_switch\_ON/OFF} (non-modifiable)
        \item \texttt{Acceleration\_speed\_-\_Longitudinal} (non-modifiable)
        \item \texttt{Short\_Term\_Fuel\_trim\_Bank1} (modifiable)
        \item \texttt{Fuel\_consumption} (modifiable)
        \item \texttt{Engine\_torque\_after\_correction} (modifiable)
        \item \texttt{Engine\_torque} (modifiable)
        \item \texttt{Current\_spark\_timing} (modifiable)
        \item \texttt{Torque\_converted\_turbine\_speed\_-\_Unfiltered} (borderline)
        \item \texttt{Torque\_converter\_speed} (borderline)
    \end{enumerate}
    } %
    \vspace{1em}
Listing 1: Top SHAP features.
    \end{minipage}
\end{figure}

Figure~\ref{fig:gancatrain} summarizes the GANCAN training pipeline discussed in Section~\ref{subsec:gan-can}.
Moreover, as discussed in Section~\ref{subsub:model}, we employed \ac{shap} to extract the most important features used by our model, which are listed in Listing 1. One key observation is that most features describe the internal engine state rather than direct driving behavior. This suggests that drivers are distinguished more by how they stress the engine, rather than by how they steer or brake. However, precise evaluation of the effect of features needs future work. 

\end{document}